\documentclass[aps,twocolumn,pra,superscriptaddress,notitlepage,floatfix]{revtex4-1}
\usepackage{amssymb, graphicx,subfigure}
\usepackage{enumerate}
\usepackage{amsmath,bm}
\usepackage{color}
\usepackage{natbib} 
\usepackage[latin1]{inputenc}

\begin{document}
\newcommand{\ket}[1]{|#1\rangle}
\newcommand{\bra}[1]{\langle #1|}
\newcommand{\escalar}[2]{\langle #1|#2\rangle}
\newcommand{\expectation}[3]{\langle #1|#2|#3\rangle}
\newcommand{\expected}[1]{\left\langle #1\right\rangle}
\renewcommand{\figurename}{\footnotesize Fig.}
\renewcommand{\tablename}{\footnotesize Table}

\title{Giant enhancement of  photodissociation of polar    
dimers in electric fields}

\author{R.\ Gonz\'alez-F\'erez}

\affiliation{Instituto Carlos I de F\'{\i}sica Te\'orica y Computacional,
and Departamento de F\'{\i}sica At\'omica, Molecular y Nuclear,
  Universidad de Granada, Spain} 

\author{P.\ Schmelcher}

\affiliation{Zentrum f\"ur Optische Quantentechnologien, Universit\"at
  Hamburg, Germany} 

\date{\today}
\begin{abstract}

We explore the photodissociation of polar dimers in static electric fields in the cold regime using the example of the LiCs
molecule.
A giant enhancement of the differential cross section is found for laboratory electric field strengths, and analyzed 
with varying rovibrational bound states, continuum energies as well as field strengths.
\end{abstract}

\maketitle

\section{Introduction}

The availability of cold to ultracold polar dimers in their  absolute 
ground state~\cite{ni08,deiglmayr:133004,PhysRevLett.105.203001}
 represents a breakthrough towards the control of all
molecular degrees of freedom, i.e., the center of mass, electronic, rotational
and vibrational motions. 
Such a polar sample provides a unique starting point  to 
manipulate 
the internal structure of molecules, and therefore the molecular dynamics with electrostatic fields.  
Oriented pendular states present a paradigm in this context. They exhibit 
a preferred spatial 
direction of their permanent electric dipole moments, which introduces an 
anisotropic and long-range electric dipole-dipole interaction 
between two dimers. 
Thus, by changing the directional features  one can control
and manipulate physical processes taking place in the
ultracold regime, e.g., scattering 
dynamics~\cite{gorshkov:073201,ticknor:133202,tscherbul:194311,avdeenkov:022707}, 
chemical reactions~\cite{Miranda:NatPhys,PhysRevA.83.012705,PhysRevA.81.060701,krems_pccp}, spin
dynamics~\cite{1367-2630-12-10-103007}, or even more, create 
quantum computational devices 
\cite{demille02,yelin06,zoller_qc}.  

Here, we investigate a specific process, namely, the dissociation of polar diatomic molecules into two cold 
atoms via single-photon absorption, 
and how this process can be controlled and manipulated by using an additional  
external electric field. 
As a  representative example, we consider  the electronic ground state of the LiCs 
dimer and analyze the direct dissociation within the electronic ground 
state.
The dependence of the angular distribution of the photofragments on the initial
rovibrational bound state, on the energy of the continuum, and on the static
electric field strength are analyzed in detail.  
Our focus is the cold regime, where transitions to a few  scattering  partial waves
contribute.
In the later case the spectrum shows shape resonances showing up in the fact that, e.g., 
an initial bound state with rotational quantum number $J\pm1$ leads to large field-free
differential cross sections and angular distributions via  a shape resonance of angular
momentum $J$.
Switching on a static field,  the hybridization of the angular motion, i.e., mixing of the rotational
states,  takes places, and the
dissociation process is significantly modified. 
We demonstrate that a moderate static electric field  provokes 
enhancements of several orders of magnitude of the angular distributions 
of the dissociation process.

The paper is organized as follows. In Sec. \ref{se:hamil} the rotational
Hamiltonian is presented together with the differential
cross section and angular distribution of a direct dissociation process within the
electronic ground state.
We discuss the numerical results for LiCs in Sec. \ref{sec:results}, analyzing 
the field-free and field-dressed cases.
The conclusions and outlook are provided in Sec. \ref{sec:conclusions}.

\section{The rovibrational Hamiltonian and the dissociation process}
\label{se:hamil}
The description of this process is achieved within the Born-Oppenheimer separation 
of the electronic and nuclear coordinates. 
We assume that for the considered regime of field strengths a 
perturbation theoretical treatment  holds for the description of the interaction of the field
with the electrons, whereas a  nonperturbative treatment is
indispensable for the corresponding nuclear dynamics.   
Thus, the rovibrational Hamiltonian of a polar dimer exposed to an homogeneous
and static electric field reads  
\begin{equation}
\label{eq:rotvib_hamiltonian}
H= T_R
+\frac{\hbar^2\mathbf{J}^2(\theta,\phi)}{2\mu R^2} + 
V(R)-FD(R)\cos\theta
\end{equation}
where $R$ and $\theta, \phi$ are the internuclear distance and the Euler
angles, respectively, and we use the molecule fixed 
frame with the coordinate origin at the center of mass of the nuclei.  
$T_R$, $\hbar\mathbf{J}(\theta,\phi)$, $\mu$, and $V (R)$ are the vibrational
kinetic energy, orbital angular momentum, reduced mass of the nuclei and the field-free
adiabatic electronic potential energy curve, respectively.
The last term in Eq.~\eqref{eq:rotvib_hamiltonian}
provides the interaction between the electric field and
the molecule via its permanent electronic dipole moment function 
$D(R)$. The electric field is  oriented along the $z$-axis of the
laboratory frame  and posses the  strength $F$. 
Our study is restricted to a non-relativistic treatment and 
addresses the spin singlet electronic ground state. 

In field-free space, each bound state of the molecule is characterized by its
vibrational, rotational, and magnetic quantum numbers $(\nu, J,M)$. In the 
presence of the electric field only the magnetic quantum
number $M$ is conserved. However, for reasons of addressability
we will refer to the electrically dressed states by
means of the corresponding field-free quantum numbers $(\nu,J,M)$.

We assume a direct dissociation process within the electronic ground state
with $^1\Sigma^+$ symmetry  caused by a beam of linearly polarized
light.
The polarization of this laser is taken parallel to the direction of the  static field. 
The initial state consists of a rovibrational
bound state of the polar dimer exposed to an electrostatic electric field  and 
via the absorption of a photon a transition takes place to a continuum final
state.
The absorption of the photon from a linear polarized laser is described in
terms of perturbation theory. Thus, making use of the dipole approximation
the differential cross section is given by
\begin{equation}
  \frac{d\sigma}{d\Omega}=\frac{\hbar^2}{2\mu}I(\Omega), 
\label{eq:diff_cross_sect}
\end{equation}
with $I(\Omega)\equiv I(\Omega;\eta,E,F)$ being the angular distribution
\begin{equation}
I(\Omega)=
|\langle\Psi_C(\Omega;E)|
D(R)\cos\theta|\Psi_{\eta}\rangle|^2 ,  
\label{eq:angular_distribution}
\end{equation}
where  $\Psi_{\eta}\equiv\Psi_{\eta}(R,\theta,\phi)$ is the rovibrational bound
eigenstate of the Hamiltonian \eqref{eq:rotvib_hamiltonian}. 
It  is characterized by the quantum numbers 
$\eta\equiv(v,J,M)$, i.e., $v$ and $J$ are the field-free
quantum numbers of this field-dressed level.
The final state $\Psi_C(\Omega;E)\equiv\Psi_C(\Omega,R,\theta,\phi;E)$
is, in the field-free case, given by~\cite{zare1972}
\begin{eqnarray}
\Psi_C(\Omega;E)=\sum_{J'M'}&&(2J'+1)e^{i\Delta_{J'M'}}
\psi_{J'M'}(R,\theta,\phi;E) \nonumber\\
&& \times D_{M'0}^{J'}(\Omega,0)
\label{eq:continuum_state}
\end{eqnarray}
where 
$\psi_{J'M'}(R,\theta,\phi;E)$ is the energy normalized eigenfunction of the
Hamiltonian \eqref{eq:rotvib_hamiltonian} describing the relative motion of
the nuclei with kinetic energy $E=\hbar k^2/2\mu>0$. It is characterized by
the phase $\delta_{J'M'}=\Delta_{J'M'}+J'\pi/2$ through the relation  
$\psi_{J'M'}(R\to\infty,\theta,\phi;E)\sim \sin(kR+\delta_{J'M'}-\pi
J'/2)Y_{J'M'}(\theta,\phi)$, where
$Y_{J'M'}(\theta,\phi)$ are the  spherical harmonics normalized-to-unity. 
The momentum $\mathbf{k}$, with Euler angles $\Omega=(\Theta,\Phi)$, 
represents the propagating vector pointing along the
final recoil direction of the photofragments.
$D_{M'0}^{J'}(\Omega,0)=D_{M'0}^{J'}(\Theta,\Phi,0)$
is a Wigner rotation matrix \cite{zare}.

We remark that if we assume the molecular ensemble to be randomly
oriented, the population is equally distributed among the states with
different values of $M$, and using the axial recoil approximation
\cite{zare1972}, the angular distribution \eqref{eq:angular_distribution} is
reduced to the well  
known expression $I(\Theta,\Phi)=\sigma(1+\beta P_2(\Theta))/4\pi$, with
$\beta$ being the anisotropy parameter, $P_2(\Theta)$ the second-order
Legendre polynomial, and it is normalized to the cross section $\sigma$. 
This expression was derived in the pioneering study of R.N. Zare and D. R.
Herschbach~\cite{zare1963}, which was followed by a vast and rich amount of 
theoretical and experimental works. In particular, several of them have been devoted to the
investigation of the photodissociation of oriented molecules
\cite{Zare19891,choi:150,wu:9447,PeterRakitzis2003187,rakitzis:science,brom:11645,VanDenBrom2006,Rakitzis2010937}

Since the probe laser is linearly polarized, we encounter in the absence of the
static electric field the selection rules are $\Delta J= \pm 1$ and $\Delta
M=0$. Thus, the angular distribution \eqref{eq:angular_distribution} is reduced
to the square of a combination of two Legendre polynomials of degree 
$J+1$ and $J-1$, and it is symmetric with respect to $\Theta=\pi/2$.
By turning on the electrostatic field, the hybridization of the rotational
motion takes place and only the selection rule $\Delta M=0$ holds. 
We restrict our analysis to initial and final states with $M=0$, and 
due to the azimuthal symmetry the angular distribution is independent of
the Euler angle $\Phi$, i.e., $I(\Omega)=I(\Theta)$.  
In our description, we consider the impact of the electric field on 
both the continuum and bound levels, and their wave functions are obtained
from solving the Schr\"odinger equation associated to the rovibrational
 Hamiltonian \eqref{eq:rotvib_hamiltonian}. 
However, due to the long-range behaviour of the dipole moment, $D(R)\to D_7 R^{-7}$, with $D_7$ being
constant, for $R\to\infty$,  
the field influence on the continuum states is smaller compared to 
the bound ones. Thus, we assume that  there is no
appreciable hybridization of its angular motion, so that the field-free
rotational quantum number can be identified, and the expression for the wave
function \eqref{eq:continuum_state}  is still holds approximately.

Our aim is to explore the properties of the angular distribution as the initial
state, the continuum energy of the photofragments, and the field strength 
are varied. 
The two-dimensional Schr\"odinger equation associated with the nuclear
Hamiltonian \eqref{eq:rotvib_hamiltonian} has  been  numerically  solved
by employing a hybrid computational method that combines discrete and
basis-set techniques applied to the radial and angular coordinates,
respectively \cite{gonzalez:023402,gonzalez04}. 
The continuum is discretized and the numerically obtained wave functions are
$L^2$-normalized. 
However, the continuum wave functions appearing in 
Eqs.~\eqref{eq:angular_distribution} and \eqref{eq:continuum_state} are
energy-normalized. We address this issue with the help of 
 a time-dependent formalism 
\cite{gonzalez:023402}. The continuum phase is computed by using the fact that
our numerical technique selects only those continuum states that are zero
on the borders of the  discretization box. 

\section{Results}
\label{sec:results}
We investigate the impact of a static electric field on the
dissociation of a polar dimer. 
As prototype example we take the $^7$Li$^{133}$Cs dimer in its electronic
ground state $X^1\Sigma^+$. 
The potential energy curve is obtained from experiment~\cite{staanum06}
including the correct long-range behavior. For
the electric dipole moment function, we have used semiempirical 
data \cite{aymar05}, which were linearly extrapolated to short
distances, and at long distance we have fitted them to the 
asymptotic behaviour $D_7/R^7$. 
Since  $D(R)$ is negative, in a strong electric field the pendular
levels are antioriented along the field direction~\cite{gonzalez06,gonzalez09,mayle06,gonzalez07_2,gonzalez:023402}.
The electronic ground state of $^7$Li$^{133}$Cs accommodates $55$ vibrational
bands, and the last one $\nu=54$ has two bound rotational levels with
$J=0$ and $1$ \cite{gonzalez09}. 
The continuum spectrum presents a shape
resonance with  energy $E\approx 1.36$~mK and rotational quantum
number $J=2$. 
Angular distribution will be provided in atomic units. 

\begin{figure}[h]
\centering
  \includegraphics[scale=.54]{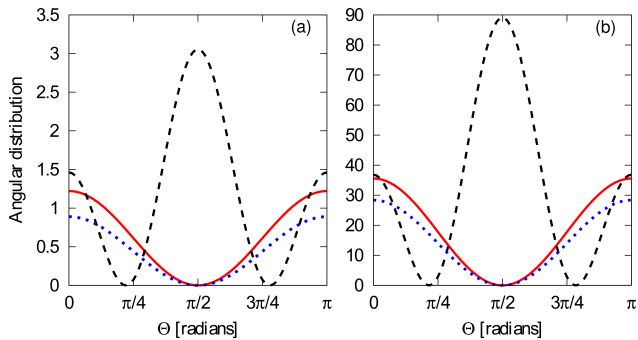}
  \caption{Field-free angular distribution for the dissociation process to a
    continuum state with energy $E\sim 0.3$~mK, the initial states
 are $(45,J,0)$ (a)  and  $(50,J,0)$(b) with rotational quantum numbers
 $J=0$ (full line),$1$ (dashed line) and $2$ (dotted line).}
  \label{fgr:fieldfree}
\end{figure}
Let us first analyze the dissociation process in the absence of the
electrostatic field.  
Figures~\ref{fgr:fieldfree}(a) and (b) present the angular distribution
as a function of the angle $\Theta$ for the initial states with 
$\nu=45$ and $50$ and $J=0,\,1$, and $2$, respectively, and the
final continuum energy is $0.3$~mK.
For a certain rotational quantum number, 
$I(\Theta)$ is symmetric with respect to $\Theta=\pi/2$ and
it is a combination of Legendre polynomials. For the $J=0$ states, the
dissociation is only possible to continuum levels with $J=1$,  and $I(\Theta)$
is proportional to $\cos^2\Theta$. 
For the $J=1$ initial states, the matrix elements of the $p\,\to\, d$-wave
transitions are larger than the corresponding ones of the $p\,\to\, s$-wave
transitions. 
In this case, $I(\Theta)$ does not have zeros but two minima 
at small but nonzero values
which 
are shifted with respect to the two nodes of the second order Legendre
polynomial. 
From $J=2$ bound states, the final continuum
level could have $J=1$ and $J=3$, with the transition to $J=1$-levels being 
dominant, and $I(\Theta)$ is zero for $\Theta=\pi/2$. 
The quantitatively different values of $I(\Theta)$ in both panels
illustrate its strong dependence on the initial bound state
due to the overlap of the bound state wave function with the highly
oscillating continuum wave function weighted by the dipole operator.   
For lower lying rovibrational levels, the differential
cross section is therefore significantly reduced.  

\begin{figure}[h]
\centering
 \includegraphics[scale=.54]{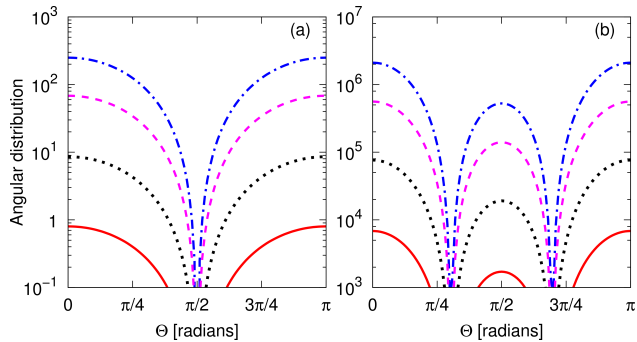}
  \caption{Field-free angular distribution for the dissociation process to a
    continuum state with energy $E\sim 1.36$~mK, which is part of the spectrum where the
    $J=2$ shape resonance appears. The initial states
 are $(\nu,0,0)$ (a)  and  $(\nu,1,0)$ (b) with vibrational quantum numbers
 $\nu=40$ (full line), $45$ (dotted line), $50$ (dash-dotted line), and
$52$ (dashed line).}
  \label{fgr:fieldfree_SR}
\end{figure}
The effect of increasing the dissociation energy is manifested in the
interplay between the allowed transitions, and, thus, in a variation of
$I(\Theta)$.
In Figs.~\ref{fgr:fieldfree_SR}(a) and (b) we present the angular distribution
from bound states with $J=0$ and $1$, respectively, within the vibrational
bands $v=40,\,45,\, 50$ and $52$, for a continuum energy in the vicinity of
 the shape
resonance. Note, that a semilogarithmic scale is used to cover the wide range
of variation of $I(\Theta)$. 
Taking the bound state belonging to the last vibrational bands 
gives rise to variations of $I(\Theta)$ of more than two orders of
magnitude. 
If  the continuum energy is increased $I(\Theta)$   is also increased, 
as can be seen from a comparison of the
results for the bound levels $(40,0,0)$ and $(50,0,0)$ presented 
in  Figs.~\ref{fgr:fieldfree}(a) and (b)
and those in 
Fig.~\ref{fgr:fieldfree_SR}(a). 
Due to the field-free selection rules, the transition to the shape
resonance is not allowed for the $(\nu,0,0)$ states, whereas
it has a dramatic impact on the dissociation
from  the $(\nu,1,0)$ levels, cf. 
Fig.~\ref{fgr:fieldfree_SR}(b). Indeed, $I(\Theta)$ shows the same dependence on $\Theta$ as 
in Figs.~\ref{fgr:fieldfree}(a) and (b), but its absolute value is
increased by several orders of magnitude. 
For the $(50,1,0)$ state, we find $I(0)=36.8$ and $2.1\times 10^6$
for continuum energies $0.3$ and $1.36$~mK, respectively.
The resonance wave function has a larger probability at short
distances, which gives rise to a larger overlap with the bound state wave
function.  
In the absence of the field, this resonance level is also accessible from
a   bound $f$-wave, and the corresponding angular distribution 
is  also significantly increased compared to other spectral regions.

On exposing the molecule to an additional static electric
field, the hybridization of the angular motion takes place, the
selection rule on $J$ does not hold any more, and the process of
dissociation is altered severely. 
We focus here on the transition from three bound rovibrational levels 
within the  field-free vibrational band $\nu=50$, and for rotational
quantum numbers $J=0,\,1,$ and $2$, with $M=0$, and 
we consider $F\le 10.28$ KV cm$^{-1}$.

\begin{figure}[h]
\centering
 \includegraphics[angle=0,scale=.75]{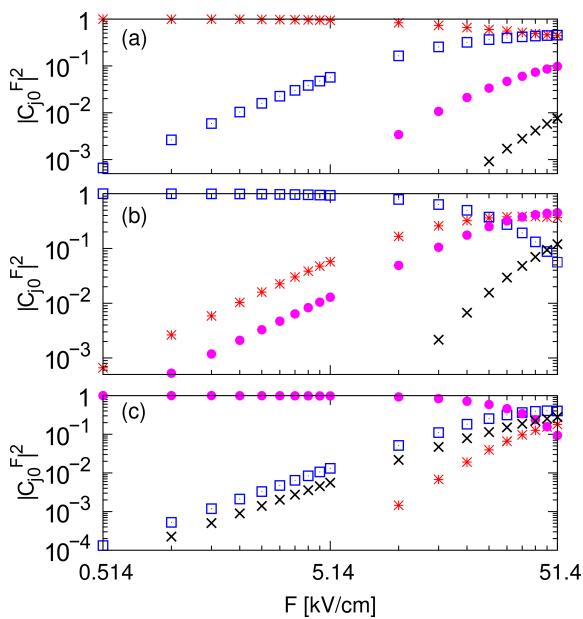}
  \caption{Contribution of the partial $s$ ($\ast$), $p$ ($\Box$), 
$d$  ($\bullet$), and $f$ ($\times$) waves
to the wavefunctions of the states
$(50,0,0)$ (a), $(50,1,0)$ (b), and $(50,2,0)$ (c) as a function
of the strength of the  static electric field.}
  \label{fgr:weigths_field}
\end{figure}

Before discussing the results for the angular distributions, it is illuminating
to analyze the evolution of the above  states as the field is varied. 
The $(50,0,0)$ level is a high-field-seeker, and 
for $F=10.28$~kV~cm$^{-1}$, it shows a moderate  orientation with 
$\langle\cos\theta\rangle=-0.451$.
In the considered regime of field strengths,
the levels $(50,1,0)$ and $(50,2,0)$  are  low-field-seekers, and 
they are not essentially oriented, $\langle\cos\theta\rangle=0.021$, and
$0.009$, respectively,  for $F=10.28$~kV~cm$^{-1}$.
Based on our previous work\cite{gonzalez09} about  highly excited
rovibrational levels of LiCs in a static electric field, we can safely assume
that the considered field strengths do not yield a significant impact on
the vibrational dynamics.
Hence, in the framework of the effective rotor
approximation \cite{gonzalez04}, for a certain field strength $F$, the
wave function of the $(v,J,M)$ state  can
be written as
\begin{equation}
\label{eq:wave_v_j}
  \Psi_{vJM}(R,\theta,\phi)\approx
  \psi_{v0,0}(R)\sum_{j=0}^{N-1}C_{jM}^FY_{jM}(\theta,\phi), 
\end{equation}
where $Y_{jM}(\theta,\phi)$ is the normalized-to-unity spherical harmonics
with indices $j$ and $M$, and $\psi_{v0,0}(R)$ are the field-free vibrational wave
function of the state $(v,0,0)$. The $C_{jM}^F$ provide the contribution of
the partial wave $(j,M)$ to the rotational dynamics. They depend on the
field strength, and satisfy the normalization condition $\sum_j|C_{jM}^F|^2=1$. 
Note that the above sum should contain an infinite number of terms, but for
reasons of 
addressability we assume that only the first $N$ partial waves are needed to
properly describe the corresponding dynamics.
The evolution  of the weights $|C_{j0}^F|^2$, $j=0,\dots,3$ for the states
$(50,J,0)$ with $J=0,\,1$ and $2$ are presented for varying $F$  in 
Figs.~\ref{fgr:weigths_field}(a), (b), and (c), respectively. 
For the $(50,J,0)$ 
level, $|C_{J0}^F|^2$ remains roughly constant and close to $1$ for $F\lesssim 10$~kV~cm$^{-1}$, decreasing thereafter. 
Moreover, its contribution is dominant for
$F\lesssim51.4,\, 25.7$ and $30.8$~kV~cm$^{-1}$ and $J=0,\, 1$ and $2$,
respectively.   
The weights of the other partial waves, $|C_{j0}^F|^2$ with $j\ne J$,
monotonically increase with increasing  $F$. 

\begin{figure}[h]
\centering
 \includegraphics[scale=.54]{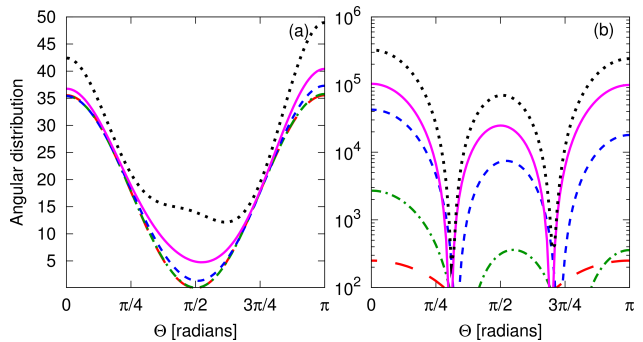}
  \caption{Field-dressed angular distribution for the dissociation process to a
    continuum state with energy $E\sim 0.3$~mK (a) and $1.36$~mK (b), 
the initial state
being $(50,0,0)$,  and  
$F=0$ (long-dashed line),
$0.514$ (dash-dotted  line), 
$2.57$  (short-dashed line), 
$5.14$  (full line), 
and $10.28$~kV~cm$^{-1}$ (dotted line).}
  \label{fgr:field_J0}
\end{figure}
\begin{figure}[h]
\centering
 \includegraphics[scale=.54]{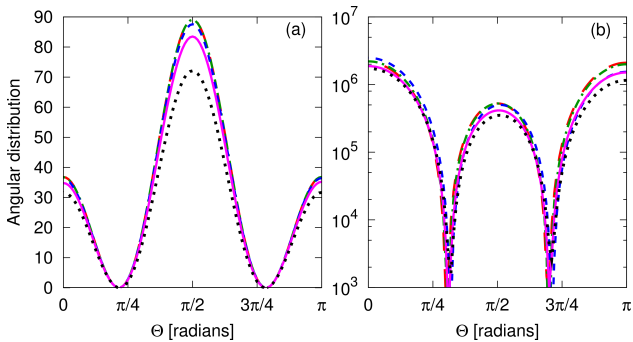}
  \caption{Same as in Fig.~\ref{fgr:field_J0} but for the initial state
$(50,1,0)$.}
 \label{fgr:field_J1}
\end{figure}
\begin{figure}[h]
\centering
 \includegraphics[scale=.54]{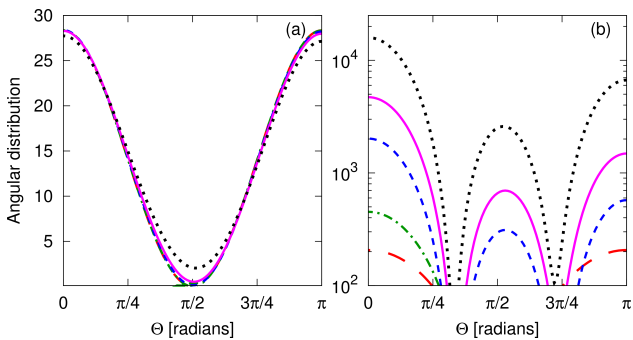}
  \caption{Same as in Fig.~\ref{fgr:field_J0} but for the initial state
$(50,2,0)$.}
  \label{fgr:field_J2}
\end{figure}

Figures~\ref{fgr:field_J0}(a, b), \ref{fgr:field_J1}(a, b), 
and \ref{fgr:field_J2}(a, b) show $I(\Theta)$  for the two 
continuum energies $0.3$ and $1.36$~mK, and for the 
 initial levels $(50,J,0)$ with
$J=0,\,1$, and $2$ as well as  field strengths 
$F=0,\,0.514,\,2.57,\,5.14$ and $10.28$~kV~cm$^{-1}$. 
For the  energy $0.3$~mK, 
the dissociation process is weakly affected by the field, 
see Figs.~\ref{fgr:field_J0}(a), 
\ref{fgr:field_J1}(a) and \ref{fgr:field_J2}(a). 
For  the initial bound state $(50,0,0)$,   
$I(\Theta)$ looses its $\cos^2\Theta$-shape as the field is increased, and 
it 
shows significantly 
larger values for $\Theta\ge\pi/2$, compared to $\Theta<\pi/2$
 for $F\ge 2.57$~kV~cm$^{-1}$,
which is due to antiorientation. In particular, 
we have
$I(0)=35.6$ and $42.4$ 
and 
$I(\pi)=35.6$ and $49.0$,  
for $F=0$ and $F=10.28$~kV~cm$^{-1}$, respectively. 
For the $(50,1,0)$ and $(50,2,0)$ states, this effect is only appreciable for 
$F=10.28$~kV~cm$^{-1}$.

If the dissociation takes place to the spectral region close to the shape
resonance, the additional static field provokes a giant enhancement of the
angular distribution of the $J=0$ and $2$ bound states, see 
Figs.~\ref{fgr:field_J0}(b), and \ref{fgr:field_J2}(b). 
For the initial state $(50,0,0)$, 
$I(\Theta)$ increases by  three orders of magnitude 
when 
the field is increased from $F=0$ to $10.28$~kV~cm$^{-1}$, 
and the corresponding value  for the angular distribution  at zero
are $I(0)=250$ to
$3.2\times 10^{5}$, respectively. 
The impact on the dissociation behaviour of the 
$(50,2,0)$ state is some what reduced:
we obtain $I(0)=207$ and $1.6 \times 10^{4}$ for 
$F=0$ and  $10.28$~kV~cm$^{-1}$, respectively.
The large enhancements of the cross sections and angular distributions are due to the
field-induced admixtures of $p$-wave ($J=1$) contributions to the $(50,0,0)$ and
$(50,2,0)$  states which couple directly to the d-wave ($J=2$) shape resonance.
Nevertheless, for $F=10.28$~kV~cm$^{-1}$, the weight of the p-wave to the
$(50,0,0)$ and $(50,2,0)$  wave functions are still relatively
small, with $|C_{10}^F|^2=0.164$ and $0.051$, respectively.
Note that these large enhancement of the angular distribution are
obtained for field strengths which are well within  experimental reach.
Furthermore, for each vibrational band, exits a field strength
that gives rise to the proper hybridization of the angular motion and 
consequently to an
enormous increase of the the dissociation probability for the rotational states
 $J\ne 1$ or $3$.
The angular distribution looses its reflection symmetry with respect to $\pi/2$
due to the large contribution of
the transition to the shape resonance. 
Analogously to the case of dissociation to the continuum with energy $E=0.3$~mK, 
the angular distribution emerging from the $(50,1,0)$ state is very
weakly affected by the external field. 
The field-dressed $I(\Theta)$ retains  its large values, and only a
minor asymmetry is observed for strong fields.


\section{Conclusions}
\label{sec:conclusions}
The dissociation of 
polar molecules in their electronic ground state by the absorption of a single photon
to produce two ground state atoms has been investigated in the presence of an
additional electrostatic field, taking as a prototype the LiCs dimer. 
We have performed a fully rovibrational description of the influence of the
electrostatic field on the molecule, whereas the interaction with the laser
field has been treated by perturbation theory. 
We concentrate on the cold regime with continuum energies around
$1$~mK, and on bound levels being vibrationally highly excited, 
$40\le\nu\le 52$, and rotationally cold, $J=0,1,2$ and $M=0$. 
Our study focused in the experimentally accessible range of field
strengths  $F=0.514-10.28$~kV~cm$^{-1}$.

In the absence of the electrostatic field, the dissociation on the cold regime 
which includes several continuum partial waves  such as $p\to s$ or
$p\to d$ transitions.   
Hence, the angular distribution of  the states $(\nu,J,0)$, with $J=0,1,$ and
$2$ shows values of the same order of magnitude.
The field-free $I(\Theta)$ is a linear combination of at most two 
Legendre  polynomials, being reflection symmetric with $\Theta$. A
strong dependence on both the bound and continuum states is encountered. 
The presence of a shape resonance gives rise to an enhancement of 
several orders of magnitude for 
the corresponding cross section of those allowed transitions.  
By turning on an electrostatic field the dissociation process is drastically altered 
due to the hybridization of the angular motion.
The transitions from the states  
emerging from the field-free $J=0$ and $2$ states to the shape resonance become
allowed and for moderate fields, the corresponding angular distribution is several 
orders of magnitude larger than its field-free
counterpart.  Specially, field strengths of a few kV~cm$^{-1}$
provoke giant enhancements of $I(\Theta)$ for the  initial states $(\nu,0,0)$,
whereas a few tens of kV~cm$^{-1}$ are needed to observe a similar impact 
on $I(\Theta)$ if the process starts from  $(\nu,2,0)$ levels.

Although our study focuses on the LiCs dimer and on the direct
dissociation taking place within its electronic ground state, we stress that
the above-observed physical phenomena are expected to occur equally
for  other
polar molecules. Shape resonances are typical features in the continuum, 
and their rotational quantum
numbers depend on the bound  states of the highly excited
vibrational bands giving plenty of freedom to the occurrence of the observed
ehancement 
effect.

\begin{acknowledgments}

We thank Enrique Buend\'{\i}a, Jes\'us S. Dehesa and Hans-Dieter Meyer, and  
for discussions. 
Financial support by the Spanish project FIS2008-02380 (MICINN) as well as the
Grants FQM-2445 and FQM-4643 (Junta de Andaluc\'{\i}a),
Campus de Excelencia Internacional Proyecto GENIL CEB09-0010  is gratefully
appreciated. R.G.F. belongs to the Andalusian research group FQM-207. 

\end{acknowledgments}

\bibliographystyle{apsrev4-1}
\bibliography{molecules} 

\end{document}